\documentclass[conference]{IEEEtran}
\IEEEoverridecommandlockouts
\usepackage{cite}
\usepackage{amsmath,amssymb,amsfonts}
\usepackage{algorithmic}
\usepackage{graphicx, import}
\usepackage{textcomp}
\usepackage{xcolor}
\usepackage{svg}
\def\BibTeX{{\rm B\kern-.05em{\sc i\kern-.025em b}\kern-.08em
    T\kern-.1667em\lower.7ex\hbox{E}\kern-.125emX}}

\newcommand\authormark[1]{\textsuperscript{#1}}

\usepackage{bm}

\usepackage{mleftright}

\usepackage{siunitx}
\usepackage{tikz, pgfplots}
\usetikzlibrary{external,plotmarks}
\usetikzlibrary{backgrounds}
\usetikzlibrary{arrows,calc,fit,matrix,positioning,shapes,shadows,trees,mindmap,tikzmark,arrows.meta,angles,quotes,babel}
\tikzstyle{every picture}+=[remember picture]
\tikzset{mark options={solid,{fill=white}}}
\pgfplotsset{compat=newest} 
\usetikzlibrary{matrix,decorations.pathreplacing}

\definecolor{ColFIR}{rgb}{0,0.4470,0.7410}
\definecolor{ColCNN}{rgb}{0.8500,0.3250,0.0980}
\definecolor{ColCNNtanh}{rgb}{0.4940,0.1840,0.5560}
\definecolor{ColGRU}{rgb}{0.4660,0.6740,0.1880}

\definecolor{ALUColor1}{rgb}{0,0.4470,0.7410}
\definecolor{ALUColor2}{rgb}{0.8500,0.3250,0.0980}
\definecolor{ALUColor3}{rgb}{0.9290,0.6940,0.1250}
\definecolor{ALUColor4}{rgb}{0.4940,0.1840,0.5560}
\definecolor{ALUColor5}{rgb}{0.4660,0.6740,0.1880}
\definecolor{ALUColor6}{rgb}{0.3010,0.7450,0.9330}
\definecolor{ALUColor7}{rgb}{0.6350,0.0780,0.1840}
\definecolor{KITgreen}{rgb}{0,.59,.51}
\definecolor{KITblue}{rgb}{.27,.39,.66}

\def\thus{\relax
	\ifmmode
		\implies
	\else
		$\implies$
	\fi}

\def\real{{\mathrm{Re}}}

\def\rz{\ifmmode{\mathds{R}}%
    \else{\hbox{$\mathds{R}$}}\fi} 
\def\nz{\ifmmode{\mathbb{N}}%
    \else{\hbox{$\mathds{N}$}}\fi} 
\def\gz{\ifmmode{\mathds{Z}}%
   \else{\hbox{$\mathds{Z}$}}\fi} 
\def\cz{\ifmmode{\mathds{C}}
    \else{\hbox{$\mathds{C}$}}\fi}%
\def\qz{\ifmmode{\mathds{Q}}%
    \else{\hbox{$\mathds{Q}$}}\fi}%
\def\K{\ifmmode{\mathds{K}}%
    \else{\hbox{$\mathds{K}$}}\fi}%

\def\Er{\ifmmode{\mathbb{E}}%
    \else{\hbox{$\mathbb{E}$}}\fi}%
    
\def\V{
	\ifmmode
	{\mathrm{V}}
	\else
	${\mathrm{V}}$
	\fi}

\usepackage[nohyperlinks,nolist]{acronym}
\acrodef{AWGN}[AWGN]{additive white Gaussian noise}
\acrodef{ANN}[ANN]{artificial neural network}
\acrodef{ADC}[ADC]{analog-to-digital converter}

\acrodef{B2B}[B2B]{back-to-back}
\acrodef{BER}[BER]{bit error rate}
\acrodef{BMI}[BMI]{binary mutual information}
\acrodef{BPS}[BPS]{blind phase search}
\acrodef{bram}[BRAM]{block random access memory}

\acrodef{CD}[CD]{chromatic dispersion}
\acrodef{CM}[CM]{constant-modulus}
\acrodef{CMA}[CMA]{constant-modulus algorithm}
\acrodef{CNN}[CNN]{convolutional neural network}
\acrodef{CPE}[CPE]{carrier-phase estimation}
\acrodef{CNN}[CNN]{convolutional neural network}

\acrodef{DAC}[DAC]{digital-to-analog converter}
\acrodef{DD}[DD]{decision-directed}
\acrodef{DFE}[DFE]{decision feedback equalizer}
\acrodef{DP}[DP]{dual-polarization}
\acrodef{DFB}[DFB]{distributed-feedback}
\acrodef{DOP}[DOP]{degree of paralellism}
\acrodef{DSP}[DSP]{digital signal processing}

\acrodef{EAM}[EAM]{electro-absorption modulator}
\acrodef{ELU}[ELU]{exponential linear unit}
\acrodef{ELBO}[ELBO]{evidence lower bound}

\acrodef{FEC}[FEC]{forward error correction}
\acrodef{FIR}[FIR]{finite impulse response}
\acrodef{FPGA}[FPGA]{field-programmable gate array}
\acrodef{FTTH}[FTTH]{fiber-to-the-home}
\acrodef{FF}[FF]{flip-flop}
\acrodef{FPGA}[FPGA]{field programmable gate array}

\acrodef{GRU}[GRU]{gated recurrent unit}

\acrodef{IM/DD}[IM/DD]{intensity-modulation and direct-detection}
\acrodef{IP}[IR]{impulse response}
\acrodef{ISI}[ISI]{inter-symbol interference}

\acrodef{KL}[KL]{Kullback-Leibler}

\acrodef{LDPC}[LDPC]{low-density parity-check}
\acrodef{LLR}[LLR]{log-likelihood ratio}
\acrodef{LMS}[LMS]{least mean squares}
\acrodef{LUT}[LUT]{look-up table}

\acrodef{MA}[MA]{moving average}
\acrodef{MAC}[MAC]{multiply-accumulate}
\acrodef{MAP}[MAP]{maximum a~posteriori}
\acrodef{MIMO}[MIMO]{multiple-input multiple-output}
\acrodef{ML}[ML]{maximum likelihood}
\acrodef{MMA}[MMA]{multi-modulus algorithm}
\acrodef{MMSE}[MMSE]{minimum mean squared error}
\acrodef{MSE}[MSE]{mean squared error}

\acrodef{NN}[NN]{neural network}
\acrodef{NRZ}[NRZ]{non-return-to-zero}

\acrodef{OOK}[OOK]{on-off-keying}
\acrodef{ONU}[ONU]{optical network unit}
\acrodef{OLT}[OLT]{optical line terminal}

\acrodef{PAM2}[PAM2]{two-level pulse amplitude modulation}
\acrodef{PAM4}[PAM4]{4-ary pulse amplitude modulation}
\acrodef{PCS}[PCS]{probabilistic constellation shaping}
\acrodef{pdf}[pdf]{probability density function}
\acrodef{PMD}[PMD]{polarization mode dispersion}
\acrodef{pmf}[pmf]{probability mass function}
\acrodef{PON}[PON]{passive optical network}
\acrodef{PSK}[PSK]{phase shift keying}
\acrodef{PSP}[PSP]{principal state of polarization}

\acrodef{QAM}[QAM]{quadrature amplitude modulation}
\acrodef{QoS}[QoS]{quality of service}

\acrodef{ReLU}[ReLU]{rectified linear unit}
\acrodef{RDE}[RDE]{radius-directed equalizer}
\acrodef{ROP}[ROP]{received optical power}
\acrodef{RRC}[RRC]{root-raised cosine}
\acrodef{rvm}[rvm]{real-valued multiplication}

\acrodef{SER}[SER]{symbol error rate}
\acrodef{SSMF}[SSMF]{standard single-mode fiber}
\acrodef{SNR}[SNR]{signal-to-noise ratio}
\acrodef{SOA}[SOA]{semiconductor optical amplifier}
\acrodef{sps}[sps]{samples per symbol}

\acrodef{VAE}[VAE]{variational autoencoder}
\acrodef{VOA}[VOA]{variable optical attenuator}
\acrodef{VQVAE}[VQVAE]{vector-quantized variational autoencoder}

\begin{document}

\title{Real-Time FPGA Demonstrator of ANN-Based Equalization for Optical Communications\\

\thanks{This work was carried out in the frameworks of the CELTIC-NEXT project AI-NET-ANTILLAS (C2019/3-3, grant 16KIS1316 and 16KIS1317), funded by the German Federal Ministry of Education and Research (BMBF), and the RENEW project, funded by the European Research Council (ERC) under the European Union's Horizon 2020 research and innovation programme (grant agreement No. 101001899). }
}


\author{
\IEEEauthorblockN{Jonas Ney,\authormark{1,*} Patrick Matalla,\authormark{2} Vincent Lauinger,\authormark{3} Laurent Schmalen,\authormark{3} Sebastian Randel,\authormark{2} and Norbert Wehn\authormark{1}}\\
\IEEEauthorblockA{
\authormark{1}Microelectronic Systems Design (EMS), RPTU Kaiserslautern-Landau, 67653 Kaiserslautern, Germany\\
\authormark{2}Institute of Photonics and Quantum Electronics (IPQ), Karlsruhe Institute of Technology (KIT), 76131 Karlsruhe, Germany\\
\authormark{3}Communications Engineering Lab (CEL), Karlsruhe Institute of Technology (KIT), 76131 Karlsruhe, Germany\\
\authormark{*}\texttt{jonas.ney@rptu.de}
}
}


\maketitle

\begin{abstract}
In this work, we present a high-throughput \ac*{FPGA} demonstrator of an \ac*{ANN}-based equalizer. The equalization is performed and illustrated in real-time for a 30 GBd, \ac*{PAM2} optical communication system.
\end{abstract}


\section{Introduction}
Short-reach optical communication systems must be cost and power efficient while satisfying the increasing demand of
data rates. For these reasons, they typically rely on \ac{IM/DD} of the optical signal. In such systems, the achievable data rate is limited due to severe nonlinear signal impairments caused for example by low-cost components such as \acp{EAM} and \acp{SOA}, or by nonlinear channel effects caused, e.g., by  \ac{CD} with square-law detection. 

Digital linear equalization can compensate for such nonlinear distortions only to a certain extent. Thus, nonlinear equalization methods draw the attention of the community. In particular, machine learning using \acp{ANN}, which proved to be highly capable of compensating for nonlinear impairments, is a contender for future equalizers~\cite{Khan2019}. However, compared to conventional algorithms, which have been optimized for decades, \acp{ANN} often introduce high computational complexity, limiting the achievable throughput of the hardware implementation. Previous works showed that even with advanced \ac{FPGA} platforms, it is challenging to meet the strict performance requirements of optical communication systems~\cite{Freire2022, Kaneda2022}.

In this work, we present an \ac{FPGA} demonstrator that tackles the challenges of real-time \ac{ANN}-based equalization by featuring an optimized \ac{ANN} topology and a highly parallelized hardware architecture. The optical \ac{PAM2} signal is transmitted at \SI{30}{GBd} over a \SI{20.56}{km} long \ac{SSMF}. The digital signal processing of the receiver, including clock recovery, \ac{ANN}-based equalization, downsampling, decisioning, and \ac{BER} calculation, is fully implemented on an \ac{FPGA}. Furthermore, the demonstrator provides a visualization of the equalization performance on a monitor in real-time.

\section{Scientific and Technical Description}

In the following, we provide a short overview of our work, starting from the algorithm and its implementation down to the actual demonstrator. For the interested reader, we refer to the provided references for a more detailed explanation.

\subsection{ANN Topology}
In contrast to classical linear equalizers based on \ac{FIR} filters, the demonstrator features an \ac{ANN}-based equalizer. Specifically, we choose a \ac{CNN} as neural network type since it resembles the structure of conventional digital filters. In particular, the model consists of \num{3}~convolutional layers with a kernel size of~\num{9} where the first two layers are followed by batch-normalization and ReLU activation function. The feature maps of the first two layers consist of \num{5}~channels while the last feature map has \num{8}~channels. Correspondingly, the first layer has a stride of~\num{8}, which allows for the processing of \num{8}~symbols with one run of the network. 

The model has been selected based on an extensive design space exploration, described in detail in~\cite{neySAMOS}.
As a result, the \ac{ANN} achieves around one order of magnitude lower \ac{BER} than a linear digital filter with the same computational complexity.

\subsection{FPGA ANN Implementation}
For \ac{FPGA} implementation of the \ac{ANN}, \textit{Vivado HLS} in combination with \textit{Vivado 2022.2} is used. The hardware architecture is based on arbitrary precision fixed-point format for weights and activations while the 6-bit inputs and outputs are transferred using the AXI-Stream interface. 

The main target of our demonstrator is to satisfy the high-throughput constraints of the \SI{30}{\giga Bd} optical communication channel. Thus, the aim of our hardware architecture is to utilize the available resources efficiently by boosting parallelism on all implementation levels. 
The first level of parallelism is based on our streaming hardware architecture, where each of the layers is implemented as an individual hardware instance. This way, the data is processed in a pipelined fashion, with each layer corresponding to a separate pipeline stage. Hence, each layer can start its operation as soon as the first inputs are received. Further, our convolutional layer exploits multiple types of parallelism: on the level of input channels, on the level of output channels, and on the kernel level. Another level of parallelism that is exploited by our implementation is the number of \ac{CNN} instances. We place and connect multiple instances of the \ac{CNN} in one design to further boost the throughput. Therefore, the input is split into multiple streams, so each instance operates on a subset of the input sequence and produces a subset of the output sequence. For further information about the hardware architecture and the quantization, we refer to~\cite{neySAMOS}.


\subsection{Demonstrator}
\begin{figure}[t]
    \centering
    \begin{tikzpicture}
[block/.style={draw,align=center,rounded corners = 0.05cm, minimum width=1cm,minimum height=0.5cm},
block2/.style={draw,align=center,rounded corners = 0.05cm, minimum width=2cm,minimum height=0.4cm}]
\definecolor{myblue}{HTML}{0072BD}%

\footnotesize 

    \node[block] at (0,0) (fpga1) {FPGA};
    \node[align=center] at (-1.25,0) {\SI{30}{GBd},\\PAM2};

    \node[block] at (0,-0.75) (dac) {DAC};
    \draw[-latex] (fpga1) -- (dac);
    
    \node[block,color=myblue] at (0,-1.75) (eam) {EAM};
    \node[align=center,color=myblue] at (-1.25,-2.2) {\SI{1540}{nm}};
    \draw[-latex] (dac) -- (eam);

    \node[block,color=myblue] at (-1.25,-1.75) (dfb) {DFB};
    \draw[-latex,color=myblue] (dfb) -- (eam);

    \draw [rounded corners=0.05cm,color=myblue] (3.5,-2) rectangle (4,-1.5);
    \draw[color=myblue] (3.85,-1.925) -- (3.85,-1.575);
    \draw[fill=white,color=myblue] (3.85,-1.645) -- (3.75,-1.825) -- (3.95,-1.825) --  cycle;
    \draw[color=myblue] (3.75,-1.645) -- (3.95,-1.645);
    \draw[-latex,color=myblue] (3.55,-1.775) -- (3.7,-1.675);
    \draw[-latex,color=myblue] (3.55,-1.925) -- (3.7,-1.825);

    \draw[-latex,color=myblue] (eam) -- (3.5,-1.75);
    \draw[fill=white,draw=myblue] (1.9,-1.45) circle (0.3);
    \draw[fill=white,draw=myblue] (2,-1.45) circle (0.3);
    \draw[fill=white,draw=myblue] (2.1,-1.45) circle (0.3);
    \node[align=center,color=myblue] at (2,-2) {\SI{20.56}{km} SSMF};

    \node[block] at (3.75,-0.75) (adc) {ADC};
    \draw[-latex] (3.75,-1.5) -- (adc);

    \draw[fill=white] (3.625,-1.4) -- (3.875,-1.4) -- (3.75,-1.2) --  cycle;

    \node[block] at (3.75,0) (fpga2) {FPGA};
    \draw[-latex] (adc) -- (fpga2);

    \node[block2] at (5.75,0.3) (dsp1) {Clock Recovery};
    \node[block2] at (5.75,-0.4) (dsp2) {Neural Network};
    \node[block2] at (5.75,-1.1) (dsp4) {Decisioning}; 
    \node[block2] at (5.75,-1.8) (dsp5) {BER Tester};

    \draw[-latex] (dsp1) -- (dsp2);
    \draw[-latex] (dsp2) -- (dsp4);
    \draw[-latex] (dsp4) -- (dsp5);

    \draw [rounded corners=0.05cm, dotted] (4.60,-2.13) rectangle (6.90,0.63);
    \draw[dotted] (4.25,0.25) -- (4.60,0.63);
    \draw[dotted] (4.25,-0.25) -- (4.60,-2.13);

\end{tikzpicture}	
    \caption{Transmission setup including digital signal processing at receiver side.}    
    \label{fig:experimental_setup}
    \vspace*{-3mm}
\end{figure}
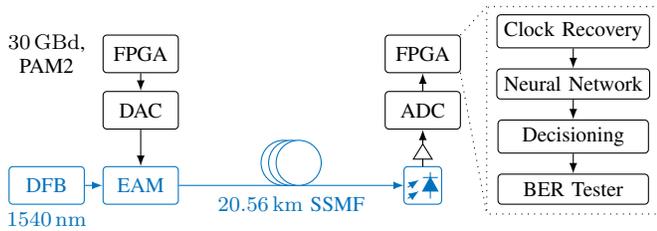

The experimental setup including the digital signal processing of the receiver is depicted in Fig.~\ref{fig:experimental_setup}. We use the Keysight USPA prototyping platform, which contains a Xilinx VU9P \ac{FPGA} on transmitter and receiver side, and a \SI{60}{GSa/s} \ac{DAC} and \ac{ADC} with \SI{6}{bit} resolution. At the transmitter side, a \SI{30}{GBd}, \ac{NRZ}, \ac{PAM2} signal is generated and drives the \ac{EAM} to modulate the optical carrier at \SI{1540}{nm}, which is provided by an \ac{DFB} laser. The optical signal is then transmitted over a \SI{20.56}{km} \ac{SSMF} to accumulate \ac{CD}. At the receiver side, the signal is detected by a photodiode and amplified by a transimpedance amplifier. Afterwards, the received signal is sampled by an \ac{ADC} and processed by the receiver \ac{FPGA}. As part of the digital signal processing of the receiver, the twofold oversampled received signal is first synchronized to the transmit clock utilizing a feedforward clock recovery architecture as described in~\cite{Matalla}. Afterwards, the signal is equalized using our high-throughput \ac{ANN}-based equalizer. Finally, we implement the decisioning and \ac{BER} tester to evaluate the signal quality and \ac{BER}. This way, the communication performance is visualized in real-time and the equalizer's compensation capabilities are showcased. 

\begin{figure}[htbp]
\def\svgwidth{\columnwidth}
\footnotesize
	\centering
        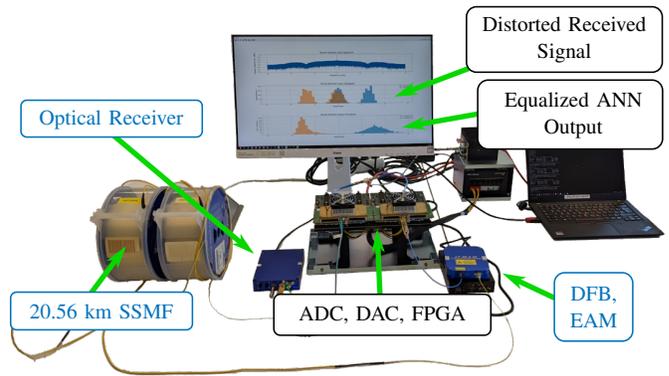
    \vspace*{-5mm}
	\caption{Setup of the real-time optical communication demonstrator.}
	\label{fig:demonstrator}
    \vspace*{-3mm}
\end{figure}

Figure~\ref{fig:demonstrator} shows the real-time demonstrator including a monitor which illustrates the functionality of the \ac{ANN}-based equalizer. Before the equalizer, the two levels of the \ac{PAM2}-symbols are partly overlapping around the decision threshold due to \ac{ISI}, which is caused by \ac{CD}, and nonlinear distortions. After the equalizer, the symbols are well separated leading to a \ac{BER} close to zero. 

Given the impracticality of transporting the costly and heavy equipment of the demonstrator, we plan to show a live stream of our demonstrator placed in our laboratory at the conference venue. On-site we will require two monitors, one for the live stream and another one for the signal and \ac{BER} monitoring. 

\section{Research Contributions \& Conclusion}

The main purpose of our demonstrator is to illustrate how \ac{ANN}-based equalizers can be applied to high-throughput optical communication systems. Thus, the demonstrator forms the basis for the visualization of multiple contributions of previous and future works in the field of \ac{ANN}-based equalization. It can be utilized to visualize the contributions of~\cite{neySAMOS}, where a similar high-throughput equalizer was implemented on the Xilinx VU13P \ac{FPGA}. Another use-case is the visualization of the novel unsupervised equalization approaches of~\cite{Ney2023} and~\cite{lauinger2024fullyblind}. Therefore, in future work, the demonstrator could further be extended to illustrate the training process of the \ac{ANN}.

To summarize, the demonstrator is a first step towards a practical optical communication system featuring \ac{ANN}-based equalizers. It shows that, if efficiently implemented, \ac{ANN}-based algorithms can satisfy the requirements of high-throughput communication systems while successfully compensating for nonlinear distortions. 

\bibliography{main}

\end{document}